\documentclass[amsmath,amssymb,aps,pra,twocolumn,floatfix,showpacs]{revtex4-1}

\usepackage{graphicx}
\usepackage{dcolumn}
\usepackage{bm}
\usepackage{subfigure}
\usepackage{epstopdf}
\usepackage{enumerate}

\newcommand{\be}{\begin{equation}}
\newcommand{\ee}{\end{equation}}
\newcommand{\ba}{\begin{array}}
\newcommand{\ea}{\end{array}}

\pdfpageattr {/Group << /S /Transparency /I true /CS /DeviceRGB>>}

\begin{document}

\title{Resonant high harmonic generation in a ballistic graphene transistor with an AC driven gate}

\author{Y. Korniyenko}
\author{O. Shevtsov}
\author{T. L\"ofwander}

\affiliation{Department of Microtechnology and Nanoscience - MC2,
Chalmers University of Technology, SE-412 96 G\"oteborg, Sweden}

\date{\today}

\begin{abstract}
We report a theoretical study of time-dependent transport in a ballistic graphene field effect transistor.
We develop a model based on Floquet theory describing Dirac electron transmission through a harmonically driven potential barrier.
Photon-assisted tunneling results in excitation of quasibound states at the barrier.
Under resonance condition, the excitation of the quasibound states leads to promotion of higher-order sidebands
and enhanced higher harmonics of the source-drain conductance.
The resonances in the main transmission channel are of the Fano form, while they are of the Breit-Wigner form for sidebands.
We discuss the possibility of utilizing the resonances in prospective ballistic high-frequency devices,
in particular frequency multipliers.
\end{abstract}


\maketitle

Already in the early years of graphene research, analogue high-frequency electronics was recognized
as a potential niche for applications \cite{Schwiertz2010,Palacios2010,GlazovPR2014,roadmap}.
Although many devices have probably been limited by parasitics due to problems with
developing good recipes for making graphene transistors,
the current speed record \cite{Cheng2012} is already a cut-off frequency of over 400 GHz.
At the same time, we have seen a rapid improvement of graphene material quality.
Mobilities reaching $10^5$ cm$^2$/V$\cdot$s at room temperature and larger than $10^6$ cm$^2$/V$\cdot$s at low temperature
have been achieved \cite{roadmap}.
Promising paths towards improved mobility include encapsulation of graphene between
layers of other two-dimensional (2D) crystals,
notably hexagonal boron-nitride, or suspension of graphene between contacts.
Using the latter approach, ultra high-quality $p-n$ junctions were recently made \cite{Rickhaus2015}.
Fabry-Perot resonances at zero magnetic field were measured, and so-called snake states were possible
to see at small magnetic fields of order 20 mT.
With such rapid improvements of device quality, it has become increasingly important to study in detail
ballistic high-frequency devices.

\begin{figure}[b]
\includegraphics[width=\columnwidth]{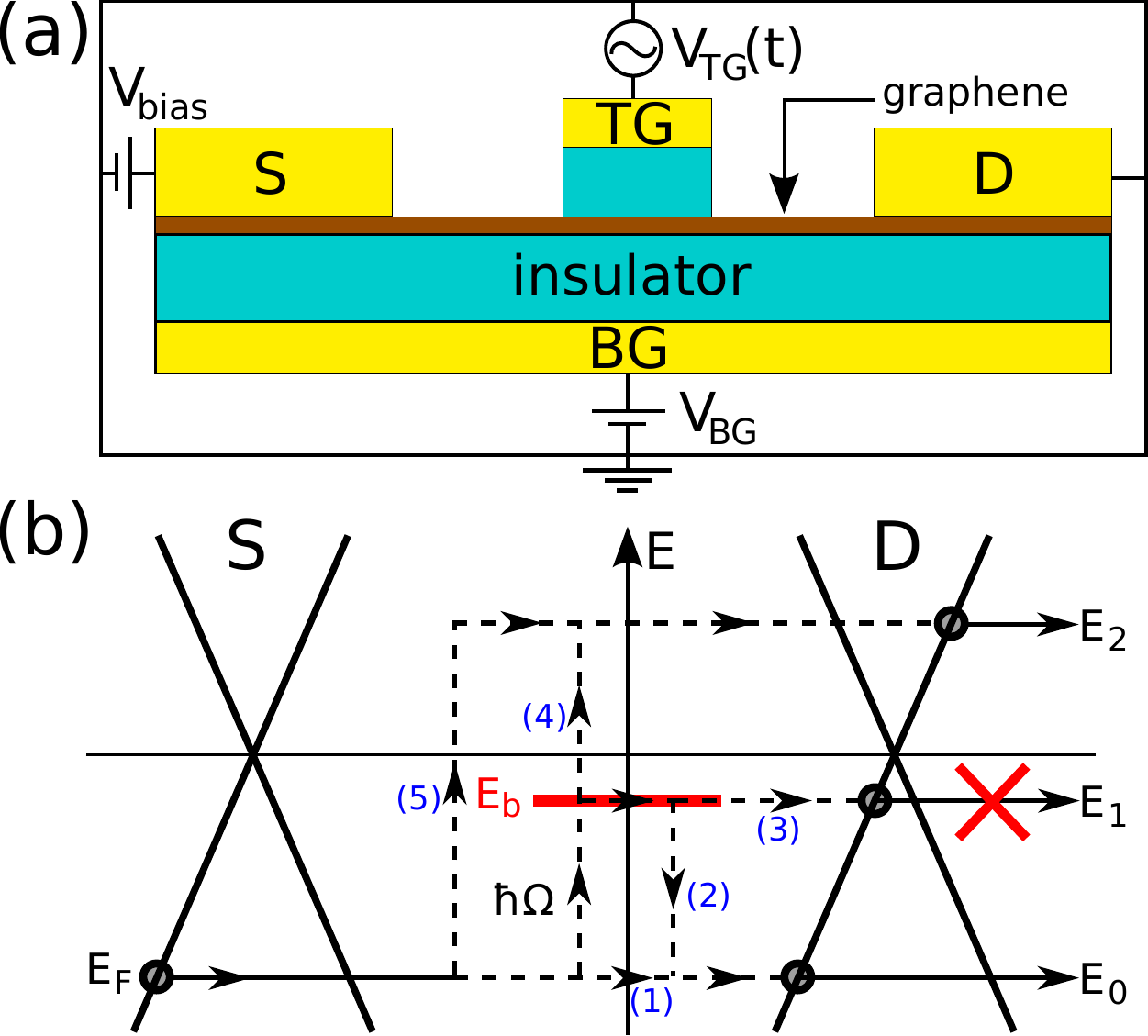}
\caption{(a) A graphene field effect transistor, where the overall doping level is controlled by
a back gate (BG), and the source (S) - drain (D) current is controlled by the top gate (TG)
dc and ac signals. (b) The harmonic ac signal of frequency $\Omega$ leads to inelastic scattering
that under resonance condition excites an otherwise unoccupied bound state in the top gate barrier potential at energy $E_b$.
This leads to a Fano resonance in transmission to $E_0$ due to interference between processes (1) and (2) and
a Breit-Wigner resonance in transmission to $E_2$ [process (4)]. Process (4) leads to higher-harmonic generation,
in particular the $2\Omega$ harmonic.}
\label{fig1}
\end{figure}

One of the key ideas behind using 2D materials for high-frequency electronics is the favorable scaling towards
short gate lengths without so-called short-channel effects \cite{Schwiertz2010}.
Thin channels (2D is the extreme) allows for short gates, high speed, and high-density integration.
High speed, reaching THz frequencies \cite{TassinScience2013}, is the ultimate goal.
Another advantage of graphene is the possibility to tune the electron density, for instance by means of a back gate:
the Fermi energy can be tuned from the electron to the hole band (through the so-called Dirac point at charge neutrality).
Such ambipolarity is very advantageous, in that both $n$-type and $p$-type devices
can in principle be made at will across a single wafer.

A challenge is to capitalize on the unique properties of graphene and derive device functionality
directly from the fact that electrons in graphene behave like massless Dirac particles
with linear spectrum and a pseudospin degree of freedom.
Several works in this direction show indeed that ac transport in graphene is a rich subject. Studies include
quantum pumping \cite{PradaPRB2009,TorresAPL2011,SanJosePRB2011,SanJoseAPL2012,Connolly2013}, 
non-linear electromagnetic response \cite{MikhailovPRL2007,Mikhailov2008,SyzranovPRB2008,CalvoAPL2012,AlNaibPRB2014,SinhaAPL2012},
and photon-assisted tunneling phenomena
\cite{Trauzettel2007,Zeb2008,RochaPRB2010,SavelevPRL2012,LuJAP2012,SzaboPRB2013,ZhuJAP2015}.
In theoretical investigations for low doping (Fermi energy $E_F$ close to the Dirac point) and high frequencies $\Omega$,
with $E_F$ and $\Omega$ of comparable magnitude (we put $\hbar=1$), a true quantum mechanical description becomes necessary.
For time-dependent transport in two-dimensional electron gases in semiconducting heterostructures,
displaying a quadratic dispersion relation,
photon-assisted tunneling in time-harmonic potentials is described well within a Floquet theory framework
and have been investigated for a long time \cite{PedersenPRB1998,platero2004,kohler2005}.
Here, we study theoretically a ballistic field effect transistor with a harmonic drive applied to the top gate, see Fig.~\ref{fig1}(a),
within a Floquet theory applicable to graphene.
The harmonic drive of frequency $\Omega$ supports inelastic scattering from the Fermi energy $E_F$, to sideband
energies $E_n=E_F+n\Omega$, where $n$ is an integer.
Near charge neutrality, on the scale of the drive frequency, the barrier is close to 
transparent due to Klein tunneling. At the same time, a quasibound state on the barrier can be inelastically excited
through a resonant process (supported by the harmonic drive) that interfere with direct elastic transmission.
This leads to a Fano resonance in direct transmission, as recently found numerically \cite{LuJAP2012,SzaboPRB2013,ZhuJAP2015}.
Here, we show that higher order sidebands are simultaneously resonantly enhanced,
which leads to the possibility of building a frequency multiplier based on a ballistic graphene device
that we study in detail in this paper.

We are interested in the intrinsic properties of the graphene transistor in Fig.~\ref{fig1}(a), and neglect parasitics.
This allows us to make a minimal model in terms of a Dirac Hamiltonian
\begin{equation}
\mathcal{H}=-i\sigma_x\nabla_x+\sigma_yk_y+\left[Z_0+Z_1\cos(\Omega t)\right]\delta(x),
\label{Hamiltonian}
\end{equation}
where we have set the Fermi velocity in graphene equal to unity, $v_F=1$.
The Pauli matrices are as usual denoted $\sigma_x$ and $\sigma_y$.
The top gate barrier potential is considered smooth on the atomic scale and cannot induce scattering
between the two valleys in the band structure. In the end all observables will contain an extra factor
of two to account for valley degeneracy, in addition to spin degeneracy.
At the same time, on the Dirac length scale (given by $\hbar v_F/E_F$ after reinstating $\hbar$ and $v_F$),
we consider the potential width $D$ to be small but its height $V$ to be large,
such that we can take the limits $D\rightarrow 0$ and $V\rightarrow\infty$ keeping the product $VD=Z$ constant.
The strengths of the time-independent component $Z_0$ and the time-dependent component $Z_1$ can be different.
The $\delta$-function in Eq.~(\ref{Hamiltonian}) is therefore smooth on the atomic scale but sharp on the Dirac length scale.
We consider the barrier to be translational invariant along the transverse direction, which guarantees that the corresponding
wave vector component $k_y$ is conserved. The spatial dependence then enters through the coordinate $x$ perpendicular to the barrier.
Finally, we assume homogeneous doping of the graphene sheet tuned by the back gate.

The methodology to solve the problem at hand is to first solve the scattering problem for the wavefunctions satisfying
the Dirac equation $\mathcal{H}\psi(x,k_y,t)=i\partial_t\psi(x,k_y,t)$. The solution can be collected into a unitary scattering matrix
for reflection and transmission coefficients between incoming waves at energy $E$ and scattered waves
at energies $E_n=E+n\Omega$.
For the current, the Landauer-B\"uttiker approach \cite{BlanterReview}
is used to compute the current operator in terms of creation and annihilation operators
for incoming and scattered waves, where the latter are related to the former through the scattering matrix.
A statistical average is performed to obtain the time-dependent current that depends on the occupation factors
of the source and drain leads, which are given by the Fermi function with chemical potentials shifted by the applied
voltage $eV$. Complete derivations of all formulas are given in the Supplemental Material \cite{SM}.
Below, we shall give results for conductance in linear response to the applied voltage at zero temperature.

For the scattering problem, since the Hamiltonian is periodic in time, we use a general Floquet ansatz
$\psi(x,k_y,t)=\sum_n\psi_n(x,k_y,E)\exp(-iE_nt)$.
When plugged into the Dirac equation it yields a set of differential equations
for the (formally infinitely many) sideband amplitudes $\psi_n(x,k_y,E)$.
In the following we do not write the arguments $x$, $k_y$, and $E$ in order to keep the notation compact.
The sideband amplitudes can be arranged into a vector
$\Phi=\left[...,\psi_{-1},\psi_0,\psi_1,...\right]^T$,
which then satisfies 
$\nabla_x\Phi=\check{M}_{td}\Phi,$
where
\begin{eqnarray}
\check{M}_{td} = \left[k_y\sigma_z+iE_n\sigma_x-iZ_0\sigma_x\delta(x)\right]\otimes\check{1}
-i\frac{Z_1}{2}\delta(x)\sigma_x\otimes\check{2}\nonumber
\end{eqnarray}
is a tridiagonal matrix in sideband space with
$(\check 1)_{nm}=\delta_{nm}$ and $(\check 2)_{nm}=\delta_{n,m+1}+\delta_{n,m-1}$.
After integration over $x=0$ we obtain a boundary condition
\begin{eqnarray}
\Phi(x=0^-) &=& \exp\left[iZ_0\sigma_x\otimes\check1+i\frac{Z_1}{2}\sigma_x\otimes\check2\right]\Phi(x=0^+)\nonumber\\
&\equiv& \check M\Psi(x=0^+).
\label{BoundaryCondition}
\end{eqnarray}
This boundary condition can also be derived by solving a square barrier problem first,
and in the end let $D\rightarrow 0$ and $V\rightarrow\infty$ keeping the product $VD=Z$ constant,
see also Ref.~\cite{McKellarPRA1987}.
This boundary condition gives an elegant view of scattering off a potential in graphene
in terms of pseudospin rotation. For instance, the static barrier leads to a rotation around the pseudospin $x$-axis by
an angle $-2Z_0$. If the pseudospin is aligned with $\sigma_x$, i.e. electron propagation along the $x$-axis with perpendicular 
incidence, the rotation has no effect (Klein tunneling \cite{Katsnelson2006}). For other angles, transmission is non-perfect.
The explicit formulas for the transmission amplitudes $t_n(k_y,E)$ derived from Eq.~(\ref{BoundaryCondition})
are given in Ref.~\cite{SM}.

\begin{figure}[t]
\includegraphics[width=\columnwidth]{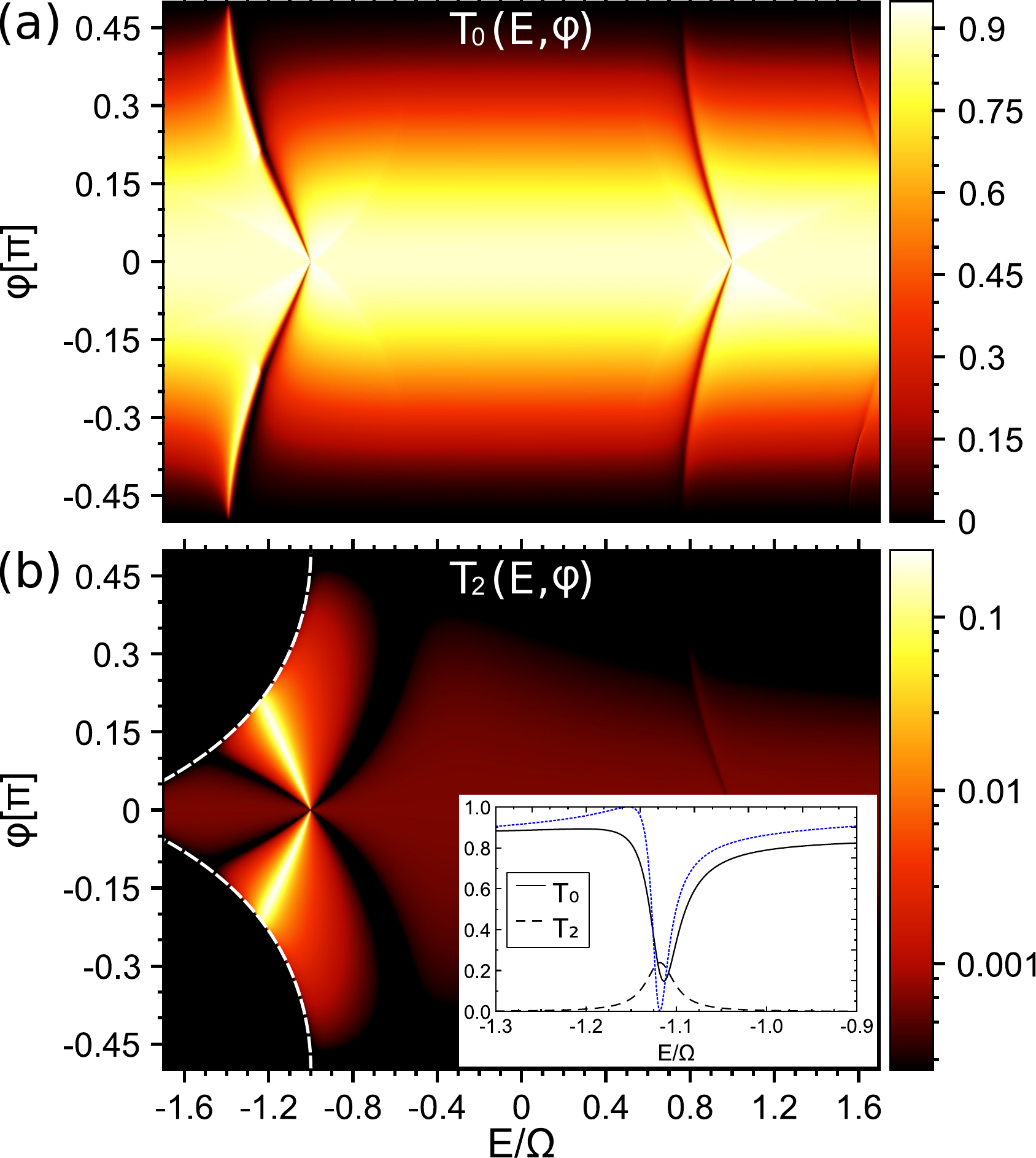}
\caption{Energy and incidence angle dependence of transmission probabilities for
(a) elastic scattering $T_0(E,\varphi)$ and (b) inelastic scattering between energy $E$ and $E+2\Omega$, $T_2(E,\varphi)$.
The black region to the left of the white dashed lines in (b) are regions where the second sideband wavefunctions
are evanescent waves decaying away from the barrier. The barrier strengths are $Z_0=0.4\pi$ and $Z_1=0.45$.
Inset: transmission probabilities for fixed $\varphi=\pi/9$.}
\label{fig2}
\end{figure}

In Fig.~\ref{fig2} we display the transmission probabilities $T_n(E,\varphi)=|t_n(E,\varphi)|^2$ for $n=0$ and $n=2$,
where $T_n(E,\varphi)$ denotes incidence on the barrier at energy $E$ and transmission at sideband energy $E_n$,
keeping the parallel momentum $k_y=|E|\sin\varphi$ conserved (the angle $\varphi$ is measured relative
to the barrier normal).
In the main transmission channel $T_0(E,\varphi)$, Klein tunneling is apparent in that the transmission is very close
to unity. Deviation from unity transmission is due to the static barrier of strength $Z_0$ and finite incidence angle (non-zero $\varphi$)
and, in addition, scattering to other sidebands with $n\neq 0$.
The transmission probability to the second sideband $T_2(E,\varphi)$ is in general very small.
For certain energies there are Fano resonances \cite{LuJAP2012,SzaboPRB2013,ZhuJAP2015}
induced by the time-dependent drive and a bound state at the barrier,
which give rise to a peak-dip structure dispersing with $\varphi$,
one feature at positive energies and another one at negative energies.
The Fano resonances occur in a parameter range where the outgoing (from the barrier) wavefunctions
at a sideband $E_{n}$ are evanescent
(below we shall concentrate on $n=\pm 1$, which are the most pronounced resonances in Fig.~\ref{fig2}).
This happens when inelastic scattering from $E=k_x(E)^2+k_y^2$ to $E_n=k_x(E_n)^2+k_y^2$ (with conserved $k_y$)
causes $k_x(E_n)$ to become imaginary.
Resonant behavior occurs due to the existence of a bound state on the barrier at
energy $E_b(Z_0,k_y)=-\mbox{sgn}(Z_0)|k_y|\cos Z_0$, see Ref.~\cite{SM},
that can be excited by the ac drive (in which case it becomes quasi-bound).
The Fano resonance at $E_r=E_b\pm\Omega$ is a quantum mechanical interference between direct
elastic tunneling and a tunneling process
involving excitation to the first sideband (for $n=\pm 1$ at $E_r=E_b\mp\Omega$) and de-excitation back
to energy $E$, see paths (1) and (2) in the diagram in Fig.~\ref{fig1}(b) for the $E<0$ case.
On resonance, inelastic tunneling to the second sideband
is resonantly enhanced and $T_{\pm 2}(E,\varphi)$ display a Breit-Wigner resonance peak at $E_r=E_b\mp\Omega$,
see Fig.~\ref{fig2}(b).
The resonance in $T_2(E,\varphi)$ can be viewed as due to transmission in energy space through a double barrier structure
with barrier heights proportional to $Z_1$.

To extract more information about the above numerical results, we proceed with an analytic analysis.
We can expand the boundary condition in Eq.~(\ref{BoundaryCondition}) to second order
in the ac drive strength $Z_1$, assuming $Z_1\ll 1$,
\begin{equation}
\check{M} \approx
e^{iZ_0\sigma_x}
\left[\check1 + \frac{iZ_1}{2}\sigma_x\otimes\check2-
\frac{Z_1^2}{8}(2\cdot\check1+\check3)\right],
\label{BCapprox}
\end{equation}
where $(\check 3)_{nm}=\delta_{n,m+2}+\delta_{n,m-2}$ in sideband space.
To second order in $Z_1$, the transmissions to the first two sidebands can be computed \cite{SM} by solving
a system of equations for $t_0$, $t_{\pm 1}$, and $t_{\pm 2}$.
We separate two cases: (i) off-resonant transmission and (ii) on-resonant transmission.
For case (i) off-resonant transmission, the equation system can be inverted directly and
we get (for each $\varphi$; we suppress the argument $\varphi$ below for brevity),
%
\begin{align}
t_0(E) &\approx \left[1 + \frac{Z_1^2}{4} + Z_1^2t^{(0)}(E)A_{0,1}(E)t^{(0)}(E_1)A_{1,0}(E)\right.\notag\\
&\left.\quad+ t^{(0)}(E)A_{0,-1}(E)t^{(0)}(E_{-1})A_{-1,0}(E)\right] t^{(0)}(E),\notag\\
t_{\pm 1}(E) &\approx -Z_1 t^{(0)}(E_{\pm 1})A_{\pm 1,0}(E)t^{(0)}(E),\notag\\
t_{\pm 2}(E) &\approx Z_1^2\left[t^{(0)}(E_{\pm 2})A_{\pm 2,\pm 1}(E)t^{(0)}(E_{\pm 1})A_{\pm 1,0}(E)\right.\notag\\
&\quad\quad\quad\left.-t^{(0)}(E_{\pm 2})A_{\pm 2,0}(E)\right]t^{(0)}(E),
\end{align}
%
where $t^{(0)}(E_n)$ is the transmission amplitude without ac drive computed at energy $E_n$,
and $A_{n,m}(E)$ is a transition amplitude in energy space between energies $E_m$ and $E_n$,
which can be related to off-diagonal matrix elements (in sideband space) of the matrix $\check M$ in Eq.~(\ref{BCapprox}).
The above expressions make the inelastic tunneling processes at play explicit, see enumerated processes in Fig.~\ref{fig1}(b).
For instance, the expression for $t_1(E)$ read from right to left has a transparent physical meaning.
It consists of transmission amplitudes at $E$ and $E_1$,
separated by a transition in energy space $A_{1,0}$, corresponding to absorption of one quantum $\Omega$.
Consequently, the process is of order $Z_1$. Direct transmission has corrections to the static transmission amplitude
due to excitation and deexcitation to neighboring sidebands (processes of order $Z_1^2$),
while $t_2(E)$ consists of a direct process of absorbing two quanta, $2\Omega$, and a sequential process involving
the first sideband energy, both are of order $Z_1^2$. This tells us that the sideband amplitudes are
in general very small when $Z_1$ is small.

The above picture changes for case (ii) on-resonant transmission, for energies near $E_r=E_b\pm\Omega$
[we shall concentrate on $E_b-\Omega$ in the following discussion, as in Fig.~\ref{fig1}(b)].
In this case, the equation determining the function $t^{(0)}(E_n,k_y)$ at energy $E_n=E_b$ (here $n=1$), has to be reconsidered.
There is a pole in the matrix equation determining the scattering matrix at this energy for fixed $k_y$, corresponding to
formation of a bound state with evanescent waves decaying away from the barrier. The bound state is unoccupied
(decoupled from reservoirs) in the absence of ac drive.
For case (ii) on-resonant transmission, we get for energies $\delta E$ around the resonance energy $E_r$
\begin{eqnarray}
t_0(E_r+\delta E) &\approx& \frac{\delta E - Z_1^2 h_2(E_r)}{\delta E + Z_1^2 h_1(E_r)} t^{(0)}(E_b),\\
t_2(E_r+\delta E) &\approx& \frac{Z_1^2 h_3(E_r)}{\delta E + Z_1^2 h_1(E_r)} t^{(0)}(E_b+2\Omega),\nonumber
\end{eqnarray}
where $h_i(E_r)\equiv h_i(E_r,k_y,Z_0)$, $i=1,2,3$, are complex functions given in Ref.~\cite{SM} (their explicit
form is not important in the discussion below). Note that $t_1$ is not
well defined near resonance (it was eliminated in the calculation) because it is related to the excitation of the bound state.
The conductance computed below will not get contributions from this sideband energy [crossed process (3) in Fig.~\ref{fig1}(b)].
For the direct transmission probability $T_0(E_r+\delta E)$,
neglecting for a while the second sideband contribution (setting $h_2=0$ above),
there is a characteristic Fano resonance form
$T_0(E_r+\delta E)\propto (q\Gamma/2+\delta E)/[\delta E^2+(\Gamma/2)^2]$,
where $\Gamma\propto Z_1^2$ and $q$ is of order unity, ${\cal O}\left[(Z_1)^0\right]$.
This is the blue dotted line displayed in the inset of Fig.~\ref{fig2}(b). Taking into account tunneling (in energy space)
to the second sideband ($h_2$ finite above) and higher, we obtain the corrected line-shape, the black solid line in
the inset of Fig.~\ref{fig2}(b).
For the probability to scatter inelastically to the second sideband, we obtain from above
a Breit-Wigner resonance with the characteristic form
$T_2(E_r+\delta E)\propto (\Gamma/2)/[\delta E^2+(\Gamma/2)^2]$, which is displayed as the 
black dashed line in the inset of Fig.~\ref{fig2}(b).
Thus, in a range $\delta E\propto Z_1^2\Omega$ around $E_r$, the response is highly non-linear
and higher-order harmonics can be resonantly enhanced.

\begin{figure}
\includegraphics[width=\columnwidth]{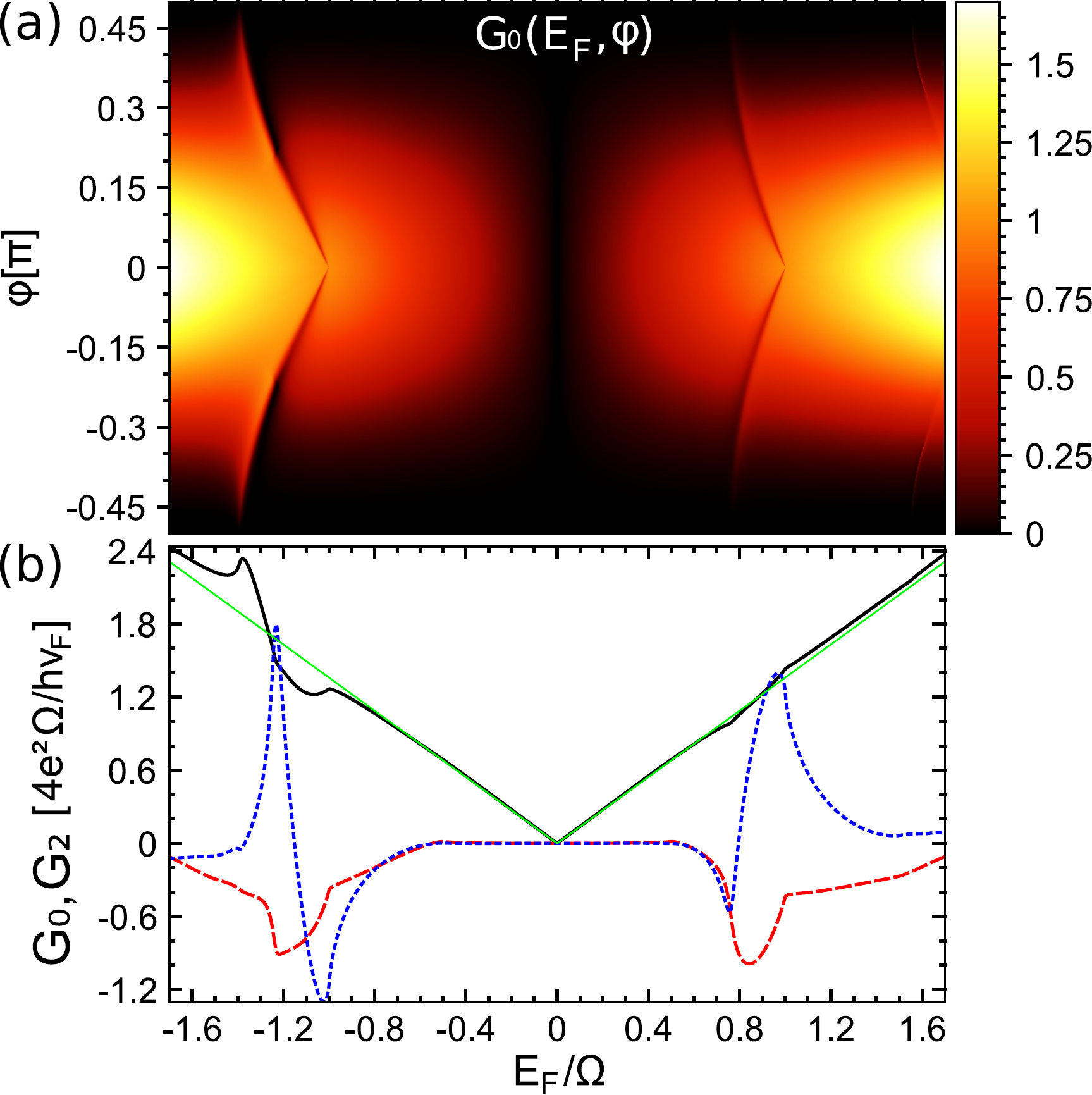}
\caption{Source-drain linear conductances in the presence of ac drive on the top gate.
Upper panel: impact angle resolved average conductance $G_0(E_F,\varphi)$.
Lower panel: angle integrated average conductance $G_0(E_F)$ and real and imaginary
parts of the second harmonic $G_2(E_F)$ (multiplied by a factor $10$).
The dip-peak structures are related to the Fano and Breit-Wigner resonances in the elastic and inelastic
transmission functions. Near resonance, the second harmonic is enhanced from ${\cal O}[Z_1^2]$ to order unity,
where $Z_1$ is the ac drive strength.}
\label{fig3}
\end{figure}

To quantify this, we present calculations of the linear conductances $G_n$, both the time-averaged component ($n=0$)
and the first harmonics ($n>0$), see Fig.~\ref{fig3}.
Note that in linear response (small source-drain voltage),
the source drain ac current $I=\sum_n I_n e^{-in\Omega t}$, with its harmonics $I_n$,
naturally define ac conductance components $G_n$, see formulas and additional figures in Ref.~\cite{SM}.
In Fig.~\ref{fig3}(a) we plot the angle resolved conductance $G_0(E_F,\varphi)$,
which reflects the sum over transmission functions in Fig.~\ref{fig2}. After angle integration,
the dc linear conductance (thin green straight lines) changes due to the ac drive into the solid black line in Fig.~\ref{fig3}(b).
The Fano resonance is clearly visible as a peak-dip feature in $G_0(E_F)$.
Thus, it is enough to study the time-averaged conductance to infer influence of the ac drive.
In Fig.~\ref{fig3}(b) we also present the real and imaginary parts of the second harmonic $G_2(E_F)$.
For small drive amplitude $Z_1$, the harmonics generally scale as $Z_1^n$ in perturbation theory and the second
harmonic is expected to be small. Near resonance, however, it is enhanced to order unity, ${\cal O}\left[(Z_1)^0\right]$, within a window
of doping $\sim Z_1^2\Omega$ around $E_F=E_r$. The results in Fig.~\ref{fig3} are obtained for a rather stronger drive $Z_1=0.45$,
including many sidebands, and the enhancement of $G_2$ is visible in a wide range of energies.
We note that there is an unfavorable prefactor for the ac components, reflected in the transmission function $T_2(E)$
being of order $0.2$ even on resonance, see Fig.~\ref{fig2}(b) and formulas in Ref.~\cite{SM}.

In summary, we have investigated time-dependent transport in a ballistic graphene field effect transistor
with ac drive on its top gate. We find resonances in inelastic scattering to sideband energies, related 
to excitation of a quasi-bound state in the top gate barrier. This leads to substantial resonant enhancement
of higher harmonics of the source-drain conductance, that could possibly be used in developing
a frequency multiplier based on a ballistic device.

\begin{acknowledgments}
It is a pleasure to thank V. S. Shumeiko for valuable discussions.
We acknowledge financial support from the Swedish foundation for strategic reseach (SSF) and
Knut and Alice Wallenberg foundation (KAW).
\end{acknowledgments}

\clearpage
\pagebreak
\widetext

\begin{center}
{\large\bf Supplemental Material}
\end{center}

\setcounter{equation}{0}
\setcounter{figure}{0}

\renewcommand{\theequation}{S\arabic{equation}}

\renewcommand{\thefigure}{S\arabic{figure}}

\section{Wave solutions in graphene}

\subsection{General solution}

We start by introducing general wave solutions in graphene without time-dependent perturbation.
They are known (see e.g. Refs.~[S\ref{S_KatsnelsonBook}-S\ref{S_FoaTorresBook}]) and we write them down here
to establish a coherent notation for subsequent sections.
As mentioned in the main text, we consider only one valley (one $K$-point) described by the Hamiltonian
\begin{equation}
\mathcal{H}_0 = -i\boldsymbol{\sigma}\cdot\boldsymbol{\nabla},\;\;\boldsymbol{\sigma}=(\sigma_x,\sigma_y).
\label{GrapheneHam}
\end{equation}
We have to solve the Dirac equation $i\partial_t\psi(x,y,t)=\mathcal{H}_0\psi(x,y,t)$, which is done by the standard ansatz
\begin{equation}
\psi(x,y,t)\propto e^{ik_xx}e^{ik_yy}e^{-iEt}\psi(k_x,k_y,E).
\end{equation}
We obtain the following eigenvalues and eigensolutions,
\begin{align}
& E_{\lambda}(k_x,k_y) = \lambda\sqrt{k_x^2+k_y^2},\;\;\lambda = \pm 1,\label{Gr_spectr}\\
& \psi_{\lambda}(k_x,k_y,E) = \frac{1}{\sqrt{2}}
\begin{pmatrix}
1\\
\frac{k_x+ik_y}{E_{\lambda}}
\end{pmatrix}
.\label{Gr_sol_1}
\end{align}

\subsection{Scattering basis}

Note that once we have found the spectrum, Eq.~(\ref{Gr_spectr}), there are only two independent parameters labeling eigenstates,
e.g. $(k_x,k_y)$ or $(k_y,E)$. Since we are going to build a scattering theory following B\"{u}ttiker [S\ref{S_Buttiker:92}-S\ref{S_BlanterReview}] the latter choice is natural because we assume translational invariance along the barrier ($y$-axis), c.f Eq.~(1) in the main text. In order to introduce the scattering basis we have to find the group velocity of states propagating along the $x$-axis (perpendicular to the barrier). Using standard definitions we have,
\begin{equation}
v(k_y,E)=\frac{\partial E}{\partial k_x}=\pm\mathrm{v}(k_y,E),\;\;\mathrm{v}(k_y,E) = \frac{\kappa_x(k_y,E)}{E},\;\;\kappa_x(k_y,E)=\mathrm{sgn}(E)\sqrt{E^2-k_y^2},
\end{equation}
where the upper and lower signs describe particles moving in the positive and negative directions along $x$, respectively.
Then we can introduce a scattering basis via
\begin{align}
&\psi_{\rightarrow}(x,k_y,E) = \frac{1}{\sqrt{2\mathrm{v}(k_y,E)}}
\begin{pmatrix}
1\\
\eta(k_y,E)
\end{pmatrix}
e^{i\kappa_x(k_y,E)x},\notag\\
&\psi_{\leftarrow}(x,k_y,E) = \frac{1}{\sqrt{2\mathrm{v}(k_y,E)}}
\begin{pmatrix}
1\\
\bar{\eta}(k_y,E)
\end{pmatrix}
e^{-i\kappa_x(k_y,E)x},\label{ScatBasis}\\
&\eta(k_y,E) = \frac{\kappa_x(k_y,E)+ik_y}{E},\;\;\bar{\eta}(k_y,E) = \frac{-\kappa_x(k_y,E)+ik_y}{E}, \notag
\end{align}
where arrows indicate the direction of propagation. The normalization in Eq.(\ref{ScatBasis}) is chosen such that a particle incident from the \textit{left} carries unit flux, defined as
\begin{equation}
j_x(x,k_y,E) = \psi^{\dagger}(x,k_y,E)\sigma_x\psi(x,k_y,E).
\end{equation}
This basis is used to find a scattering matrix and build the scattering field theory below.

\section{Floquet scattering matrix in graphene with AC $\delta$-potential}

Let us now discuss the Floquet scattering matrix for graphene in presence of an oscillating line scatterer,
i.e. we consider a system described by [c.f. Eq.~(1) in the main text]
\begin{equation}
\mathcal{H} = \mathcal{H}_0 + \left[Z_0+Z_1\cos(\Omega t)\right]\delta(x). \label{FullHam}
\end{equation}
Before discussing the solution associated to the full time-dependent Hamiltonian, it is instructive to consider $Z_1=0$.

\subsection{Static $\delta$-barrier}

In this case, since scattering is elastic, it is easy to write down a scattering ansatz, assuming incoming particles from the \textit{left},
\begin{align}
\psi(x,k_y,E) = 
\begin{cases}
\psi_{\rightarrow}(x,k_y,E)+r^{(0)}\psi_{\leftarrow}(x,k_y,E)& \text{if } x<0,\\
t^{(0)}\psi_{\rightarrow}(x,k_y,E)& \text{if } x>0.
\end{cases}
\end{align}
The superscript $X^{(0)}$ indicates functions $X$ computed for a static barrier.
The unknown transmission $t^{(0)}(k_y,E)$ and reflection $r^{(0)}(k_y,E)$ coefficients
are found through the boundary condition at $x=0$, which reads [c.f. Eq.(3) in the main text]
\begin{equation}
\psi(0^{-},k_y,E) = \exp[iZ_0\sigma_x]\psi(0^{+},k_y,E).\label{BC_static}
\end{equation}
It is straightforward to find a solution to Eq.(\ref{BC_static}), but it is convenient for what follows
to write down an equation satisfied by $t^{(0)}$,
\begin{gather}
D(k_y,E)t^{(0)} = 1,\notag\\
D(k_y,E) = \frac{1}{2\mathrm{v}(k_y,E)}
\begin{pmatrix}
-\bar{\eta}(k_y,E) & 1
\end{pmatrix}
\exp[iZ_0\sigma_x]
\begin{pmatrix}
1 \\ \eta(k_y,E)
\end{pmatrix}
.\label{D-function}
\end{gather}
Using Eq.(\ref{ScatBasis}) we can easily simplify Eq.(\ref{D-function}) and obtain
\begin{equation}
t^{(0)}(k_y,E) = D(k_y,E)^{-1}  = \left(\cos Z_0 + i\frac{\sin Z_0}{\mathrm{v}(k_y,E)}\right)^{-1}. \label{t_0}
\end{equation}
If we introduce an incidence angle $\varphi$ via $k_y = |E|\sin\varphi$, then Eq.(\ref{t_0}) can be rewritten as
\begin{equation}
t^{(0)}(k_y,E) = \frac{\cos\varphi}{\cos\varphi\cos Z_0 + i\sin Z_0}. \label{t_0_phi}
\end{equation}

\subsection{Barrier-induced bound state}\label{SecBS}

It is well-known [S\ref{S_ScatTheorBook}] that poles of the scattering matrix correspond to bound states. In our case the static $\delta$-barrier induces exactly one bound state as will be shown now. We equate to zero the denominator of Eq.(\ref{t_0}) and impose a condition that the bound state solution has to be decaying away from the barrier, which means
\begin{align}
\kappa_x(k_y,E) = i\sqrt{k_y^2-E^2},\;\;\kappa_x(k_y,E) = -iE\tan Z_0. \label{BoundStateCond}
\end{align}
One can see that Eq.(\ref{BoundStateCond}) is periodic in $Z_0$ and we consider for definiteness $-\frac{\pi}{2}<Z_0<\frac{\pi}{2}$. Then the energy of the bound state is given by
\begin{equation}
E_{b} = -\mathrm{sgn}(Z_0)|k_y|\cos Z_0.
\end{equation}
It is interesting to note that the bound state plays no role in DC transport since it is disconnected from the continuum of propagating waves. This circumstance changes as soon as we allow inelastic scattering on the barrier, when $Z_1\neq 0$.

\subsection{Oscillating $\delta$-barrier}

In the case when $Z_1\neq 0$ the Hamiltonian, Eq.~(\ref{FullHam}), is periodic in time, which enables us to use the Floquet theorem
[S\ref{S_PedersenPRB1998},S\ref{S_platero2004},S\ref{S_kohler2005}] for finding eigenvectors,
\begin{equation}
\psi(x,k_y,t) = e^{-iEt}\sum_{n=-\infty}^{+\infty}e^{-in\Omega t}\psi_n(x,k_y,E).
\end{equation}
Now if we introduce a column vector
\begin{equation}
\Phi(x,k_y,E) = \left(\dots\psi_{-1}(x,k_y,E), \psi_0(x,k_y,E), \psi_{1}(x,k_y,E),\dots\right)^{\mathrm{T}},
\end{equation}
then the condition to be satisfied at $x=0$ is [see also Eq.(3) of the main text],
\begin{align}
&\Phi(0^{-},k_y,E) = \check{M}\Phi(0^{+},k_y,E),\notag\\
&\check{M} = \exp\left[iZ_0\sigma_x\otimes\check1+i\frac{Z_1}{2}\sigma_x\otimes\check2\right],\\
&\left[\check1\right]_{n,m} = \delta_{n,m},\;\;\left[\check2\right]_{n,m} = \delta_{n,m+1}+\delta_{n,m-1}.\notag
\end{align}
Since the barrier is active only at $x=0$, asymptotic solutions are still given by a linear combination of the static solutions, Eq.(\ref{ScatBasis}). The barrier only scatters an incident particle with quantum numbers $(E,k_y)$ into a linear combination of states with quantum numbers $(E_n,k_y)$, where $E_n = E+n\Omega$ (in the end we have to consider only propagating outgoing waves, $E_n>|k_y|$, for calculating transport properties). Therefore we use the following ansatz
\begin{align}
\psi_n(x,k_y,E) = 
\begin{cases}
\delta_{n,0}\psi_{\rightarrow}(x,k_y,E_n)+r_n\psi_{\leftarrow}(x,k_y,E_n)& \text{if } x<0,\\
t_n\psi_{\rightarrow}(x,k_y,E_n)& \text{if } x>0.
\end{cases}
\end{align}
We can eliminate reflection coefficients $r_n(k_y,E)$ and find a system of equations for $t_n(k_y,E)$ only, which reads
\begin{align}
\sum_{m}\frac{1}{2\sqrt{\mathrm{v}(k_y,E_n)\mathrm{v}(k_y,E_m)}}
\begin{pmatrix}
-\bar{\eta}(k_y,E_n) & 1
\end{pmatrix}
\left[\check{M}\right]_{nm}
\begin{pmatrix}
1 \\ \eta(k_y,E_m)
\end{pmatrix}
t_m = \delta_{n,0}.\label{ACSysEqs}
\end{align}
Eq.~(\ref{ACSysEqs}) must be solved numerically.

\section{Analysis of side-band transmission coefficients: Fano and Breit-Wigner resonances}

\begin{figure}[t]
\includegraphics[width=\textwidth]{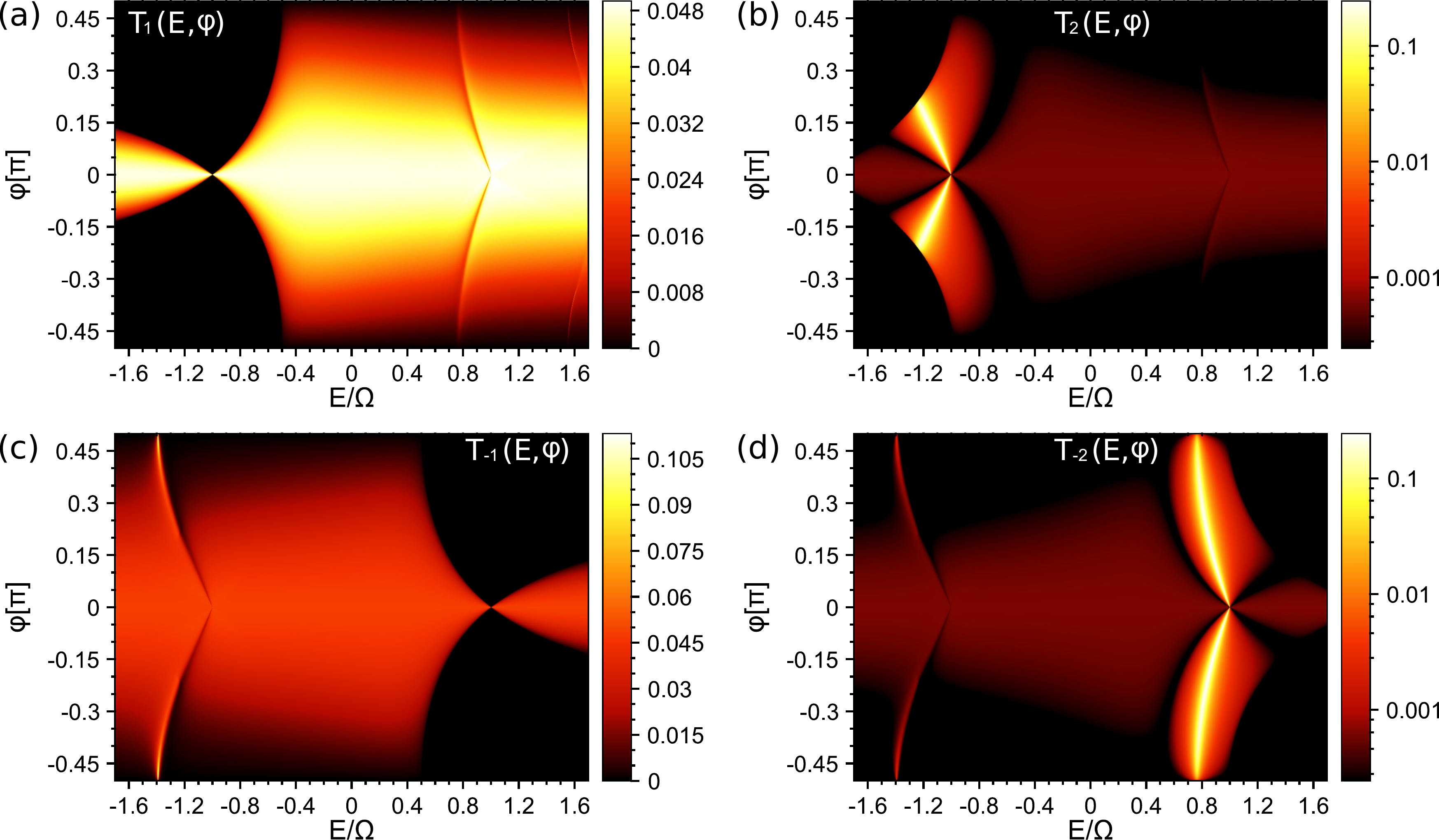}
\caption{Energy and incidence angle dependence of transmission probabilities
for inelastic scattering to the side-bands.}
\label{suppl_trans}
\end{figure}

The system of equations (\ref{ACSysEqs}) is in principle infinite in sideband index space. To find an approximate solution we have to cut the system by setting a maximum allowed $n_{\mathrm{max}}$ number of side-band. We assume that $Z_1\ll 1$, expand $\check{M}$ up to terms of order $O(Z_1^2)$, and consider five outgoing channels with $n=\left\{0, \pm 1, \pm 2\right\}$. Then we obtain a system of five coupled equations which reads [omitting the arguments $(k_y,E)$ for brevity]
\begin{align}
\begin{cases}
\left(1-\frac{Z_1^2}{4}\right)D_2t_2+Z_1A_{2,1}t_1+Z_1^2A_{2,0}t_0 = 0,\\
Z_1A_{1,2}t_2+\left(1-\frac{Z_1^2}{4}\right)D_1t_1+Z_1A_{1,0}t_0+Z_1^2A_{1,-1}t_{-1} = 0,\\
Z_1^2A_{0,2}t_2+Z_1A_{0,1}t_1+\left(1-\frac{Z_1^2}{4}\right)D_0t_0+Z_1A_{0,-1}t_{-1}+Z_1^2A_{0,-2}t_{-2} = 1,\\
Z_1^2A_{-1,1}t_1+Z_1A_{-1,0}t_0+\left(1-\frac{Z_1^2}{4}\right)D_{-1}t_{-1}+Z_1A_{-1,-2}t_{-2} = 0,\\
Z_1^2A_{-2,0}t_0+Z_1A_{-2,-1}t_{-1}+\left(1-\frac{Z_1^2}{4}\right)D_{-2}t_{-2} = 0,
\end{cases}
\label{ACSysEqs_appr}
\end{align}
where we have used the following notations
\begin{gather}
D_n(k_y,E) = D(k_y,E_n),\notag\\
A_{n,m} = \frac{(i/2)^{|n-m|}}{|n-m|}\frac{1}{2\sqrt{\mathrm{v}(k_y,E_n)\mathrm{v}(k_y,E_m)}}
\begin{pmatrix}
-\bar{\eta}(k_y,E_n) & 1
\end{pmatrix}
\exp[iZ_0\sigma_x]\sigma_x^{|n-m|}
\begin{pmatrix}
1 \\ \eta(k_y,E_m)
\end{pmatrix},
\quad n\neq m.
\label{A-functions}
\end{gather}
Note that from Eq.~(\ref{A-functions}) and Eq.~(\ref{D-function}) it is obvious that $D_n^{-1}(k_y,E)\equiv t^{(0)}(k_y,E_n)$ provided the corresponding wave is propagating, i.e. $E_n>|k_y|$. On the other hand the new functions $A_{n,m}$ have a meaning of transition matrix  between the side-bands. Now we recall that the presence of a (static) $\delta$-barrier implies existence of a bound state, see Sec.\ref{SecBS}, which now can be coupled to the propagating waves via inelastic scattering. In this case one of the functions $D_n(k_y,E)$ vanishes when $E_n=E_b$. This possibility leads to resonances in the transmission spectrum of the side-bands [see Fig.\ref{suppl_trans}], as will be discussed in details below.

\subsection{Off-resonant transmission}

We will first consider the rather trivial case of transmission in different side-bands away from the resonances. In this case we can straightforwardly estimate orders of magnitude for the side-band transmission coefficients keeping only contributions $O(Z_1^2)$,
\begin{align}
& t_0 = t_0^{(0)}+Z_1^2\tau_0,\notag\\
& t_{\pm 1} = Z_1\tau_{\pm 1},\label{OffRes_ts}\\
& t_{\pm 2} = Z_1^2\tau_{\pm 2},\notag
\end{align}
where we introduced for convenience $t_n^{(0)}(k_y,E)\equiv t^{(0)}(k_y,E_n)$. Keeping the same order of approximation in Eq.(\ref{ACSysEqs_appr}) we can easily solve it with the following result
\begin{align}
&\tau_{0} = \left(\frac{1}{4}+t_0^{(0)}A_{0,1}\frac{A_{1,0}}{D_{1}}+t_0^{(0)}A_{0,-1}\frac{A_{-1,0}}{D_{-1}}\right)t_0^{(0)},\notag\\
&\tau_{\pm1} = - \frac{A_{\pm1,0}}{D_{\pm1}}t_0^{(0)},\label{OffRes_taus}\\
&\tau_{\pm2} = \left(\frac{A_{\pm2,\pm1}}{D_{\pm2}}\frac{A_{\pm1,0}}{D_{\pm1}}-\frac{A_{\pm2,0}}{D_{\pm2}}\right)t_0^{(0)},\notag
\end{align}
which were also collected into Eq.~(4) in the main text.
The transmission coefficients, Eq.(\ref{OffRes_ts}), supplemented by Eq.(\ref{OffRes_taus}) have a physically transparent form if they describe propagating waves, i.e. waves with all $E_n>|k_y|$. In this case we can identify $D^{-1}_{n}=t_n^{(0)}$ [see Eq.(\ref{D-function})] and, reading the resulting expressions from right-to-left, we can distinguish the transmission processes depicted in Fig.1(b) in the main text [except that $E_1\neq E_b$ in the process (3), according to our assumption].

\subsection{Close-to-resonance transmission}

Now we will focus on the resonances associated with the case when the energy of one of the $n=\pm 1$ side-bands hits the bound state, $E_{\pm1} = E_b$, and the corresponding channel is closed. They are observed as zeros in $T_{\pm1}$ and maxima in $T_{\pm 2}$, dispersing with the incidence angle $\varphi$ (see Fig.~\ref{suppl_trans}). For definiteness we will consider the resonance condition for $n=1$, but this analysis is straightforward to repeat for $n=-1$. So, the resonance condition reads
\begin{equation}
D_1(k_y,E_r) = 0.
\end{equation}
We expand the $D_1$ coefficient in Eq.(\ref{ACSysEqs_appr}) around the resonance energy assuming
\begin{align}
& E=E_b-\Omega+\delta E=E_r+\delta E,\;\;|\delta E|\ll \{\Omega, |k_y|\},\notag\\
& D_1(k_y,E)\approx\frac{\delta E}{|k_y|\sin^2Z_0}.
\end{align}
Evaluating all other functions in Eqs.~(\ref{ACSysEqs_appr}) at $E=E_r$,
we solve the resulting system of equations keeping only terms of order $O(\delta E, Z_1^2)$. 
The solution for $t_0$ and $t_2$ reads
\begin{align}
& t_0 = \frac{\delta ED_2-Z_1^2A_{1,2}A_{2,1}|k_y|\sin^2Z_0}{\delta E D_0 D_2-Z_1^2\left(D_2A_{0,1}A_{1,0}+D_0A_{1,2}A_{2,1}\right)|k_y|\sin^2Z_0}, \label{t_0_appr}\\
& t_2 = \frac{Z_1^2A_{2,1}A_{1,0}|k_y|\sin^2Z_0}{\delta E D_0 D_2-Z_1^2\left(D_2A_{0,1}A_{1,0}+D_0A_{1,2}A_{2,1}\right)|k_y|\sin^2Z_0}.\label{t_2_appr}
\end{align}
We note that for $|k_y|\rightarrow 0$, $|t_0|^2$ will be close to unity due to Klein tunneling [S\ref{S_Katsnelson:06}] and there is no resonance behavior. If we consider the case when both the main channel $n=0$ and the second side-band $n=2$ are propagating, then Eqs.(\ref{t_0_appr})-(\ref{t_2_appr}) can be rewritten as
\begin{align}
& t_0 = \frac{\delta E-Z_1^2A_{1,2}t_2^{(0)}A_{2,1}|k_y|\sin^2Z_0}{\delta E-Z_1^2\left(t_0^{(0)}A_{0,1}A_{1,0}+A_{1,2}t_2^{(0)}A_{2,1}\right)|k_y|\sin^2Z_0}t_0^{(0)}, \label{t_0_appr2}\\
& t_2 = \frac{Z_1^2A_{2,1}A_{1,0}t_0^{(0)}|k_y|\sin^2Z_0}{\delta E-Z_1^2\left(t_0^{(0)}A_{0,1}A_{1,0}+A_{1,2}t_2^{(0)}A_{2,1}\right)|k_y|\sin^2Z_0}t_2^{(0)},\label{t_2_appr2}
\end{align}
with the short hand notation $t_0^{(0)}=t^{(0)}(E_r)$ and $t_2^{(0)}=t^{(0)}(E_r+2\Omega$).
\begin{figure}[t]
\includegraphics[width=0.8\textwidth]{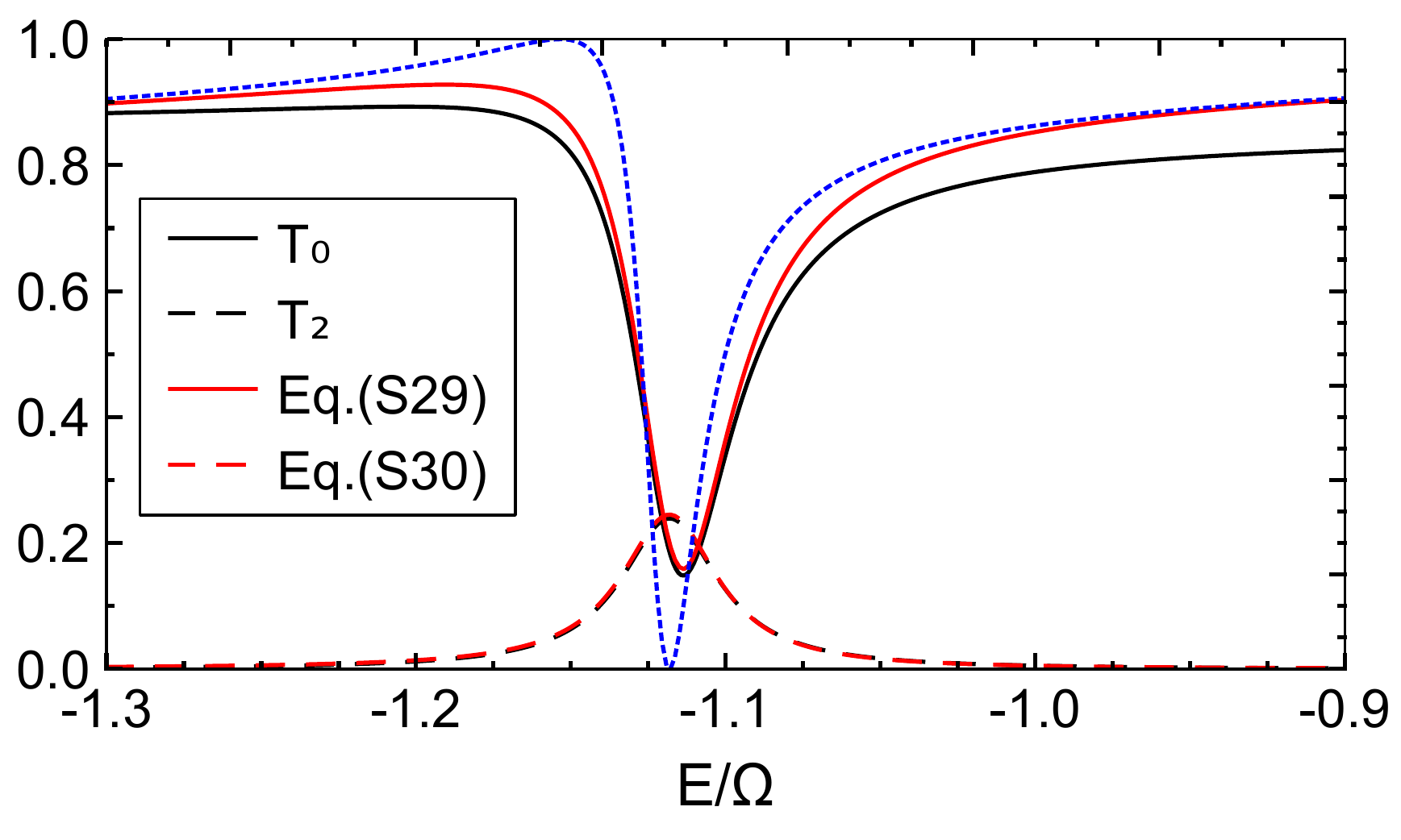}
\caption{Energy dependence of transmission probabilities $T_0$ and $T_2$ for incidence angle $\varphi = \pi/9$, $Z_0 = 0.4\pi$ and $Z_1 = 0.45$. The blue dotted line is the result for $T_0$ when neglecting scattering to the second sideband at $E_2$. In this case, the Fano resonance is fully developed (peak at unit transmission and dip at zero transmission).}
\label{appr_trans}
\end{figure}
If we analyze these expressions we can see the following:
\begin{itemize}
\item if we compute the corresponding transmission probabilities, $T_0 = |t_0|^2$, and $T_2 = |t_2|^2$, we clearly see that $T_0$ has a Fano-type resonance shape, while $T_2$ is of Breit-Wigner type with the width of the resonances $\propto Z_1^2|k_y|\sin^2 Z_0$.

\item exactly at the resonance, $\delta E\rightarrow 0$, both $t_0$ and $t_2$ have finite values independent of $Z_1$ due to constructive interference between the first and the second side-bands.
\end{itemize}
In Fig.~\ref{appr_trans} we compare the approximate solution we have found with the exact numerical calculation [see inset of Fig.2(b) in the main text]. We clearly see that Eqs.(\ref{t_0_appr2})-(\ref{t_2_appr2}) correctly describe all the essential features of the transmission probabilities discussed above.

Finally, in a more strict expansion of all functions in Eq.~(\ref{ACSysEqs_appr}) to linear order in $\delta E$, more cumbersome
expressions are obtained, but the above conclusions will not change, as also supported by the good agreement
between the black and red lines in Fig.~\ref{appr_trans}.

\section{Scattering field theory of AC current}

In this section we briefly describe the method we used to compute AC electric current. The theory below is valid as soon as a single-particle approach is justified, i.e. when particle-particle interactions can be neglected. Without loss of generality we assume particles incident on the barrier from the contact $\alpha$ [e.g. the source contact, see Fig.1(a) in the main text]. Using the scattering basis, Eq.(\ref{ScatBasis}), found above we construct a field operator
\begin{equation}
\hat{\Psi}_{\mathrm{\alpha}}(x,y,t) = \int\limits_{-\infty}^{+\infty}\frac{dk_y}{\sqrt{2\pi}}e^{ik_yy}\int\limits_{|E|>|k_y|}\frac{dE}{\sqrt{2\pi}}e^{-iEt}\left[\hat{\gamma}_{\mathrm{\alpha,in}}(k_y,E)\psi_{\rightarrow}(x,k_y,E)+\hat{\gamma}_{\mathrm{\alpha,out}}(k_y,E)\psi_{\leftarrow}(x,k_y,E)\right],
\end{equation}
in the local coordinate system of the contact, where $\hat{\gamma}_{\mathrm{\alpha,in/out}}(k_y,E)$ are the corresponding annihilation operators for the incoming/outgoing particles, which satisfy
\begin{gather}
\left\{\hat{\gamma}_{\mathrm{\alpha,in}}(k_y,E),\hat{\gamma}^{\dagger}_{\mathrm{\beta,in}}(k^{\prime}_y,E^{\prime})\right\} = \delta_{\alpha,\beta}\delta(k_y-k^{\prime}_y)\delta(E-E^{\prime}),\label{CommRels}\\
\left\{\hat{\gamma}_{\mathrm{\alpha,in}}(k_y,E),\hat{\gamma}_{\mathrm{\beta,in}}(k^{\prime}_y,E^{\prime})\right\} = \left\{\hat{\gamma}^{\dagger}_{\mathrm{\alpha,in}}(k_y,E),\hat{\gamma}^{\dagger}_{\mathrm{\beta,in}}(k^{\prime}_y,E^{\prime})\right\} = 0.\notag
\end{gather}
According to the scattering theory the outgoing operator $\hat{\gamma}_{\mathrm{\alpha,out}}(k_y,E)$ is, via a scattering matrix, related to the incoming one. For our case of an AC barrier and static contacts this relation reads
\begin{align}
\hat{\gamma}_{\mathrm{\alpha,out}}(k_y,E) = \sum\limits_{\beta}\sum\limits_{n,\mathrm{propag.}}S_{\alpha\beta}(k_y;E,E_n)\hat{\gamma}_{\mathrm{\beta,in}}(k_y,E_n),
\end{align}
where we restrict the sum over side-bands to propagating waves only, which is equivalent to setting the scattering matrix elements to zero if an incoming/outgoing wave is evanescent. Then we construct the current operator defined by the standard expression [S\ref{S_Buttiker:92}]
\begin{align}
\hat{I}_{\alpha}(x,t) = e\int\! dy\, \hat{\Psi}^{\dagger}_{\mathrm{\alpha}}(x,y,t)\sigma_x\hat{\Psi}_{\mathrm{\alpha}}(x,y,t),
\label{CurrOp}
\end{align}
where $e$ is the electron charge. Note that $\delta(k_y-k^{\prime}_y)$ in Eq.(\ref{CommRels}) must be understood in a sense of a Kronecker symbol meaning that we use Born-von Karman periodic boundary conditions in the y-direction. It means that there is a correspondence
\begin{gather}
\delta(k_y-k^{\prime}_y) = \int\limits_{-\infty}^{+\infty}\!\frac{dy}{2\pi}\,e^{i(k_y-k^{\prime}_y)y} \Leftrightarrow
\frac{1}{L_y}\int\limits_{0}^{L_y}\!dy\,e^{i(k^n_y-k^m_y)y} = \delta_{n,m},\\
\frac{2\pi}{L_y}\sum\limits_{k^n_y} \Leftrightarrow \int\limits_{-\infty}^{+\infty}\!dk_y.
\end{gather}

To obtain an observable quantity $I_{\alpha}(x,t)$ we compute a statistical average of Eq.(\ref{CurrOp}) with the help of
\begin{align}
\langle\hat{\gamma}^{\dagger}_{\mathrm{\alpha,in}}(k_y,E)\hat{\gamma}_{\mathrm{\beta,in}}(k_y,E^{\prime})\rangle 
= \delta_{\alpha,\beta}\delta(E-E^{\prime})f_{\alpha}(E),
\end{align}
where $f_{\alpha}(E)$ is a Fermi-Dirac distribution in the contact $\alpha$. The resulting expression has the form
\begin{align}
I_{\alpha}(x,t) = \sum\limits_{n=-\infty}^{+\infty}e^{-in\Omega t}I_{\alpha,n}(x),\;\;I_{\alpha,-n}(x) = I^{\ast}_{\alpha,n}(x),\label{I_gen}
\end{align}
where
\begin{align}
I_{\alpha,n}(x) = e\int\limits_{-\infty}^{+\infty}\!dk_y\int\limits_{|E|>|k_y|}\!dE\Biggl\lbrace&\delta_{n,0}f_{\alpha}(E)\notag\\
+&\frac{\eta^{\ast}(k_y,E)+\bar{\eta}(k_y,E_n)}{2\sqrt{\mathrm{v}(k_y,E)\mathrm{v}(k_y,E_n)}}e^{-i\left[\kappa_x(k_y,E)+\kappa_x(k_y,E_n)\right]x}S_{\alpha\alpha}(k_y;E_n,E)f_{\alpha}(E)\notag\\
+&\frac{\bar{\eta}^{\ast}(k_y,E_{-n})+\eta(k_y,E)}{2\sqrt{\mathrm{v}(k_y,E_{-n})\mathrm{v}(k_y,E)}}e^{i\left[\kappa_x(k_y,E_{-n})+\kappa_x(k_y,E)\right]x}\left[S_{\alpha\alpha}(k_y;E_{-n},E)\right]^{\dagger}f_{\alpha}(E)\label{I_n}\\
+&\sum\limits_{\beta}\sum\limits_{m=-\infty}^{+\infty}\frac{\bar{\eta}^{\ast}(k_y,E)+\bar{\eta}(k_y,E_n)}{2\sqrt{\mathrm{v}(k_y,E)\mathrm{v}(k_y,E_n)}}e^{i\left[\kappa_x(k_y,E)-\kappa_x(k_y,E_n)\right]x}\notag\\
&\times\left[S_{\alpha\beta}(k_y;E,E_m)\right]^{\dagger}S_{\alpha\beta}(k_y;E_n,E_m)f_{\beta}(E_m)
\Biggr\rbrace\notag.
\end{align}
Using unitarity of the scattering matrix [S\ref{S_MoskaletsButtiker_pumps}],
\begin{align}
\sum_{\alpha}\sum_{n}\left[S_{\alpha\beta}(k_y;E_n,E_m)\right]^{\dagger}S_{\alpha\gamma}(k_y;E_n,E) = \delta_{\beta,\gamma}\delta_{m,0},\\
\sum_{\beta}\sum_{n}S_{\gamma\beta}(k_y;E_m,E_n)\left[S_{\alpha\beta}(k_y;E,E_n)\right]^{\dagger} = \delta_{\alpha,\gamma}\delta_{m,0},
\end{align}
we can rewrite Eq.(\ref{I_n}) in the following form
\begin{align}
I_{\alpha,n}(x) = e\int\limits_{-\infty}^{+\infty}\!dk_y\int\limits_{|E|>|k_y|}\!dE\Biggl\lbrace
&\frac{\eta^{\ast}(k_y,E)+\bar{\eta}(k_y,E_n)}{2\sqrt{\mathrm{v}(k_y,E)\mathrm{v}(k_y,E_n)}}e^{-i\left[\kappa_x(k_y,E)+\kappa_x(k_y,E_n)\right]x}S_{\alpha\alpha}(k_y;E_n,E)f_{\alpha}(E)\notag\\
+&\frac{\bar{\eta}^{\ast}(k_y,E_{-n})+\eta(k_y,E)}{2\sqrt{\mathrm{v}(k_y,E_{-n})\mathrm{v}(k_y,E)}}e^{i\left[\kappa_x(k_y,E_{-n})+\kappa_x(k_y,E)\right]x}\left[S_{\alpha\alpha}(k_y;E_{-n},E)\right]^{\dagger}f_{\alpha}(E)\label{I_n2}\\
+&\sum\limits_{\beta}\sum\limits_{m=-\infty}^{+\infty}\frac{\bar{\eta}^{\ast}(k_y,E_m)+\bar{\eta}(k_y,E_{n+m})}{2\sqrt{\mathrm{v}(k_y,E_m)\mathrm{v}(k_y,E_{n+m})}}e^{i\left[\kappa_x(k_y,E_m)-\kappa_x(k_y,E_{n+m})\right]x}\notag\\
&\times\left[S_{\alpha\beta}(k_y;E_m,E)\right]^{\dagger}S_{\alpha\beta}(k_y;E_{n+m},E)\left[f_{\beta}(E)-f_{\alpha}(E_m)\right]
\Biggr\rbrace\notag.
\end{align}
In contrast with the usual B\"{u}ttiker theory [S\ref{S_BlanterReview}], one cannot in general neglect the energy dependence of $\kappa_x(k_y,E)$ and $\mathrm{v}(k_y,E_m)$  in Eq.(\ref{I_n2}), because the Fermi energy $E_F$ in graphene can be tuned to the Dirac point. On the other hand, if we keep the first two terms on the rhs of Eq.(\ref{I_n2}), we see that the AC current is formally determined by the full Fermi sea rather than states close to the Fermi surface only.

\section{AC differential conductance}

\begin{figure}[t]
\includegraphics[width=0.785\textwidth]{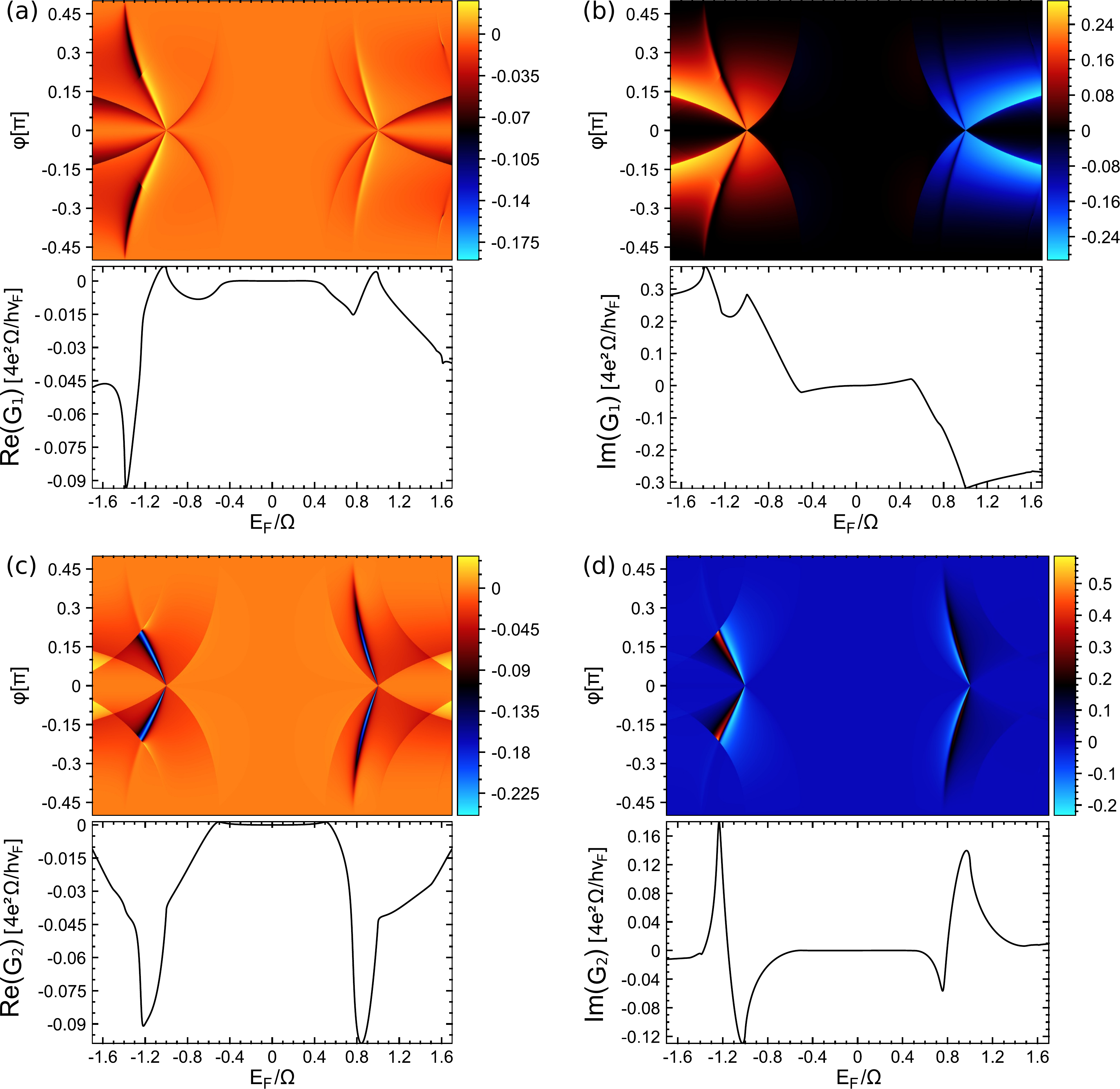}
\caption{Source-drain linear conductances for side-band currents with $n=\pm1$ and $n=\pm2$ in the presence of ac drive on the top gate.
Upper panels: impact angle resolved average conductances $G_n(E_F,\varphi)$.
Lower panels: angle integrated real and imaginary parts of average conductances $G_n(E_F)$.}
\label{suppl_cond}
\end{figure}
In this section we present formulas that we use to compute AC conductance for different side-bands in the main text. We assume that our system [see Fig.~1(a) in the main text] is at low temperature and compute a linear differential conductance with respect to the source-drain bias voltage $V_S$,
\begin{gather}
f_{\alpha}(E) = f(E-eV_{\alpha}),\;\;-\frac{\partial f(E)}{\partial E}\rightarrow\delta(E-E_F),\notag\\
G_{n}(E_F) = \left.\frac{\partial I_{D,n}(x=0^{+},V_S)}{\partial V_S}\right|_{V_S\rightarrow 0}.
\end{gather}
Note that in principle the current, Eq.~(\ref{I_gen}), is a function of coordinate and we choose the point $x=0^{+}$ in our calculations. If we use the results of the previous section we obtain
\begin{align}
G_{n}(E_F) = \frac{e^2}{h}\left.\int\limits_{-\infty}^{\infty}\!dk_y\sum_{m=-\infty}^{+\infty}\frac{\eta^{\ast}(k_y,E_m)+\eta(k_y,E_{n+m})}{2\sqrt{\mathrm{v}(k_y,E_m)\mathrm{v}(k_y,E_{n+m})}}t^{\dagger}_{m}(k_y,E)t_{n+m}(k_y,E)\right|_{E=E_F},\label{G_Ef_formula}
\end{align}
where we have restored $h$ to obtain the well-known conductance unit. This formula was used in Fig.~(3) of the main text. Finally, in Fig.~(\ref{suppl_cond}) we present the results obtained with the help of Eq.~(\ref{G_Ef_formula}) for side-bands with $n=\pm1, \pm2$.

\section*{References for Supplemental Material}

\begin{enumerate}[{[S1]}]

\item M. I. Katsnelson, {\it Graphene: carbon in two dimension}, Cambridge University Press, United Kingdom 2012.\label{S_KatsnelsonBook}

\item L. E. F. Foa Torres, S. Roche, and J.-C. Charlier, {\it Introduction to graphene-based nanomaterials}, Cambridge University Press, United Kingdom 2014.\label{S_FoaTorresBook}

\item M. B\"uttiker, Phys. Rev. B \textbf{46}, 12485 (1992).\label{S_Buttiker:92}

\item M. H. Pedersen and M. B\"uttiker, Phys. Rev. B {\bf 58}, 12993 (1998).\label{S_PedersenPRB1998}

\item Ya. M. Blanter and M. B\"uttiker, Phys. Rep. {\bf 336}, 1 (2000).\label{S_BlanterReview}

\item J. R. Taylor, \textit{Scattering Theory: The Quantum Theory on Nonrelativistic Collisions}, Dover Publications (2006).\label{S_ScatTheorBook}

\item G. Platero and R. Aguado, Phys. Rep. {\bf 395}, 1 (2004).\label{S_platero2004}

\item S. Kohler, J. Lehmann, and P. H\"anggi, Phys. Rep. {\bf 406}, 379 (2005).\label{S_kohler2005}

\item M. I. Katsnelson, K. S. Novoselov, and A. K. Geim, Nat. Phys. \textbf{2}, 620 (2006).\label{S_Katsnelson:06}

\item M. Moskalets and M. B\"{u}ttiker, Phys. Rev. B \textbf{69}, 205316 (2004).\label{S_MoskaletsButtiker_pumps}

\end{enumerate}

\end{document}